%% file: arxiv.tex
\documentclass[11pt]{article}

\usepackage[final]{acl}
\usepackage{times}
\usepackage{latexsym}
\usepackage{amssymb}
\usepackage[T1]{fontenc}
\usepackage[utf8]{inputenc}

\usepackage{microtype}
\usepackage{graphicx}
\usepackage{booktabs}
\usepackage{subcaption}
\usepackage{amsmath}
\usepackage{algorithm}
\usepackage{algpseudocode}
\usepackage{latexsym}
\usepackage{makecell}
\usepackage{textcomp}
\usepackage{tabularx} % 引入tabularx包
\usepackage{times}
\usepackage{latexsym}
\usepackage{multirow}
\usepackage{amssymb}
\usepackage{amsmath}
\usepackage{listings}
\usepackage{amsmath}
\usepackage[utf8]{inputenc}
\usepackage{float}

\usepackage{inconsolata}

\usepackage{graphicx}

\title{The WER Trap: Shattering the Illusion of Unified Tokens in Speech Language Models}

\author{Xiangyu Zhang$^{1}$, \textbf{Yuxin Li}$^{2}$, \textbf{Haoyang Zhang}$^{2}$,  \textbf{Shiqi Han}$^{1}$, \textbf{Hexin Liu}$^{2}$, \\ \textbf{Qiquan Zhang}$^{1}$, \textbf{Beena Ahmed}$^{1}$, \textbf{Julien Epps}$^{1}$\\
The University of New South Wales$^1$,
Nanyang Technological University$^2$}

\begin{document}
\maketitle
\begin{abstract}
The pursuit of a "unified" discrete token for both speech understanding and generation has led the Speech Language Model (SLM) community to heavily rely on Word Error Rate (WER)—the core metric for Whisper-style tokenizers—as the definitive proxy for representation quality. This fosters the assumption that low-WER tokens inherently preserve the information necessary for intelligible acoustic synthesis. We argue this is fundamentally deceptive. While high-frequency tokens succeed in generation tasks due to implicit information leakage, isolating pure semantic information at ultra-low frame rates strips away the fine-grained articulation and micro-dynamics essential for ODE-based generation. Empirically validating this requires extreme compression without sacrificing WER—a methodological bottleneck, as standard fixed-stride downsampling arbitrarily truncates phonetic boundaries. To overcome this, we develop a dynamic compression tokenizer that intelligently aligns representations with semantic boundaries, achieving ultra-low frame rates with exceptionally low WER. Using these isolated "pure" semantic tokens, we expose the WER trap: when conditioning generative models—even with oracle duration alignments—the reconstructed speech suffers from severe articulation blur and is rendered acoustically unintelligible. Our findings demonstrate that semantic categorization rewarded by low WER is inherently orthogonal to the continuous phonetic trajectories required for synthesis, shattering the illusion of the unified token and advocating for explicitly decoupled speech representations.
\end{abstract}

\section{Introduction}
\input{Image_set/concept}
The integration of speech modalities into large language models (LLMs) has fundamentally bifurcated the landscape of audio processing~\cite{team2023gemini,yang2025qwen3}. On the comprehension side, state-of-the-art models often leverage continuous representations from pre-trained encoders (e.g., Whisper~\cite{radford2023robust}) to maintain rich semantic density~\cite{chu2024qwen2,xu2025qwen3,wu2025step,tian2025step}. Conversely, the generative frontier has increasingly pivoted towards discrete speech tokens to enable efficient autoregressive modeling within standard NLP architectures~\cite{du2024cosyvoice,du2025cosyvoice}. This architectural divergence has fueled a persistent quest for a "unified" discrete token: a singular, highly-compressed representation capable of bridging this chasm by facilitating both robust language understanding and high-fidelity acoustic synthesis~\cite{defossez2024moshi,zhang2025mimo}. To maximize LLM efficiency—extending context windows and accelerating inference—there is an aggressive push to drive these unified tokens toward low frame rates~\cite{zhang2025mimo}.

Within this paradigm, the community has come to rely on Word Error Rate (WER)—the primary optimization objective for semantic-driven encoders—as the definitive proxy for token quality~\cite{xu2025qwen3,wu2025step}. This reliance fosters a pervasive assumption in the SLM community: that tokens optimized for minimal WER inherently preserve the essential information for intelligible acoustic generation~\cite{xu2025qwen3,wu2025step}. We argue this is fundamentally deceptive. While high-frequency discrete tokens may appear to support generation, their success is often a byproduct of implicit information leakage afforded by dense sampling. We hypothesize that as tokens are pushed toward the semantic limits required by modern SLMs, the process of isolating pure linguistic states at ultra-low frame rates inherently strips away the fine-grained articulation and micro-dynamics essential for continuous generative models.

% Empirically validating this hypothesis—that semantic purity at ultra-low frame rates is fundamentally incompatible with acoustic generation—presents a significant methodological bottleneck. To observe this generative collapse, one must achieve extreme token compression without sacrificing the downstream WER. However, standard fixed-stride downsampling mechanisms widely used in current codecs arbitrarily truncate boundaries~\cite{huang2025step,wu2025step,du2025cosyvoice}. As the compression rate increases (i.e., lowering the frame rate), this rigid downsampling aggressively corrupts the topological sequence of speech, artificially degrading the WER. Consequently, the community has been precluded from conducting a fair assessment; the degradation in generation quality at ultra-low frequencies is often mistakenly attributed to alignment errors or boundary corruption, rather than the inherent orthogonality between discriminative and generative representations.

Empirically validating this hypothesis---that semantic purity at ultra-low frame rates is fundamentally incompatible with acoustic generation---presents a significant methodological bottleneck. To observe this generative collapse, one must achieve extreme token compression without sacrificing the WER of the tokenizer itself. However, standard fixed-stride downsampling mechanisms widely used in current codecs arbitrarily truncate phonetic boundaries~\cite{huang2025step,wu2025step,du2025cosyvoice}. As the compression rate increases (i.e., lowering the frame rate), this rigid downsampling aggressively corrupts the topological sequence of speech, artificially degrading the WER~\cite{zhang2025impact,yang2025u}. Consequently, the community has been precluded from conducting a fair assessment; the degradation in generation quality at ultra-low frequencies is often mistakenly attributed to alignment errors or boundary corruption, rather than the fundamental mismatch between discriminative and generative representations.

To bypass this bottleneck and truly isolate ``pure'' semantic tokens at ultra-low frame rates, we develop a novel dynamic compression tokenizer. Rather than relying on rigid, fixed-stride time windows, our method intelligently aligns continuous representations with their underlying acoustic-semantic boundaries through a macroscopic fixed-ratio, microscopic dynamic alignment mechanism. This methodological shift allows us to achieve ultra-low frame rates while preserving exceptionally low WER, effectively providing the controlled experimental condition that has been absent from prior investigations.

With this capability, we are positioned to directly confront the central question motivating this work: \textit{does a discrete speech token that is demonstrably sufficient for language understanding---as certified by low WER and strong reasoning performance---also preserve the information necessary for high-fidelity continuous acoustic generation?} To answer this, we propose a Dual-Probing Protocol that evaluates the \textit{same} ultra-low-rate token sequence through two independent pathways: a discriminative understanding probe built on frozen LLMs, and a generative probe based on continuous Flow Matching conditioned with oracle temporal alignment. By explicitly controlling for both semantic integrity and temporal confounds, this protocol isolates the representational adequacy of the tokens themselves as the sole variable under test. Through this framework, we expose what we term the \textit{WER Trap}---demonstrating that the gap between discriminative sufficiency and generative viability is not an engineering artifact to be resolved with better alignment, but a structural consequence of extreme semantic compression---and argue that the SLM community should move toward explicitly decoupled speech representations.
% \section{Background and Problem Formulation}
% \label{sec:preliminary}

% \section{Paradigms of Speech Integration in LLMs: Continuous vs. Discrete}
\section{\textcolor{black}{Speech Representation in LLMs: Continuous vs. Discrete}}

\label{sec:preliminary}

Recent advancements in SLMs generally bifurcate into two distinct architectural paradigms regarding how audio is integrated into the text-centric space of LLMs. 

% The first paradigm, exemplified by models such as Qwen-Audio, Step-Audio, and SALMONN \cite{chu2024qwen2, tian2025step,tangsalmonn}, relies on \textbf{continuous acoustic representations}. These architectures typically employ a robust, pre-trained acoustic encoder (e.g., Whisper~\cite{radford2023robust} or Wav2Vec2~\cite{baevski2020wav2vec}) to map raw audio waveforms into a high-dimensional, continuous latent space $\mathcal{H} \in \mathbb{R}^{T \times d}$. These dense embeddings are then projected directly into the LLM's input space. While this continuous integration preserves extremely rich semantic and paralinguistic information---yielding exceptional performance on discriminative comprehension tasks \textcolor{red}{audio understanding tasks may be better as ASR may not belong to "discriminative" comprehension tasks} like ASR and Audio Question Answering (AQA)---it inherently clashes with the discrete, autoregressive generative nature of LLMs. Consequently, native end-to-end speech generation within this continuous framework is highly non-trivial, often requiring external, cascaded TTS modules or complex continuous-space diffusion adapters \cite{ho2020denoising,lipman2022flow}.

The first paradigm, exemplified by models such as Qwen-Audio, Step-Audio, and SALMONN \cite{chu2024qwen2, tian2025step,zhang2026step,tangsalmonn}, relies on \textbf{continuous acoustic representations}. These architectures typically employ a robust, pre-trained acoustic encoder (e.g., Whisper~\cite{radford2023robust} or Wav2Vec2~\cite{baevski2020wav2vec}) to map raw audio waveforms into a high-dimensional, continuous latent space $\mathcal{H} \in \mathbb{R}^{T \times d}$. These dense embeddings are then projected directly into the LLM's input space. While this continuous integration preserves extremely rich semantic and paralinguistic information---yielding exceptional performance on speech understanding tasks such as ASR and Audio Question Answering (AQA)---it inherently clashes with the discrete, autoregressive generative nature of LLMs. Consequently, native end-to-end speech generation within this continuous framework is highly non-trivial, often requiring external, cascaded TTS modules or complex continuous-space diffusion adapters \cite{ho2020denoising,lipman2022flow,yan2025step,zhang2025mamba,zhang2024speaking}.

To circumvent this generation bottleneck, a second paradigm has rapidly emerged, driving the development of fully \textbf{discrete SLMs} (e.g., Mini-Audio, SpeechGPT, and AudioLM~\cite{borsos2023audiolm,zhang2025mimo,zhang2023speechgpt,zhang2026your}). This approach employs a neural tokenizer to quantize the continuous speech signal into a sequence of discrete tokens $Z = \{z_1, z_2, \dots, z_N\}$ from a finite vocabulary $\mathcal{V}$, effectively casting audio as a ``foreign language''. This discrete formulation elegantly aligns with the standard next-token prediction objective, enabling native, autoregressive speech-to-speech and text-to-speech generation.

However, within this discrete paradigm, a further dichotomy exists regarding the nature of the tokens: acoustic versus semantic. Acoustic neural codecs (e.g., EnCodec~\cite{defossezhigh} and SoundStream~\cite{zeghidour2021soundstream}) excel at high-fidelity waveform reconstruction but suffer from high frame rates and multi-layer codebooks (e.g., Residual Vector Quantization~\cite{lee2022autoregressive}), which severely burden LLM context windows and exhibit low semantic density, hindering complex language comprehension~\cite{defossez2024moshi,zhang2023speechgpt,zhang2026duplexsla}. 

% To balance comprehension capabilities with the autoregressive efficiency of 1D sequences, the community increasingly favors \textbf{semantic-driven discrete tokens}~\cite{du2024cosyvoice,du2025cosyvoice}. Typically derived by applying quantization bottlenecks (e.g., FSQ~\cite{mentzerfinite}) to robust encoders like Whisper~\cite{radford2023robust}, these tokens are rigidly optimized for linguistic content and universally evaluated using the WER. Crucially, because these tokens empirically succeed in driving downstream continuous generative models (e.g., Flow Matching) to reconstruct high-fidelity speech at \textit{standard} frame rates, a pervasive illusion has taken root. This apparent success has led the SLM community to converge on a dangerous assumption regarding the \textbf{``unified'' discrete token}: it is widely presumed that WER serves as a universal proxy for overall token quality. Consequently, the prevailing belief is that as long as a token maintains a sufficiently low WER---even when subjected to extreme compression---it remains fully capable of facilitating both perfect language comprehension and high-fidelity continuous speech synthesis.

To balance comprehension capabilities with the autoregressive efficiency of 1D sequences, the community increasingly favors \textbf{semantic-driven discrete tokens}~\cite{du2024cosyvoice,du2025cosyvoice}. Typically derived by applying quantization bottlenecks (e.g., FSQ~\cite{mentzerfinite}) to robust encoders like Whisper~\cite{radford2023robust}, these tokens are rigidly optimized for linguistic content and universally evaluated using the WER. Crucially, because these tokens empirically succeed in driving continuous generative models (e.g., Flow Matching) to reconstruct high-fidelity speech at \textit{standard} frame
rates~\cite{du2024cosyvoice,du2025cosyvoice}, the practice of using WER as the sole proxy for overall token quality has become deeply entrenched. Recent unified SLMs routinely adopt a single set of semantic tokens---optimized and validated primarily through WER---for both language comprehension and speech synthesis~\cite{defossez2024moshi,zhang2025mimo,wu2025step}, without explicitly verifying whether the information preserved by low-WER tokens is also sufficient for continuous acoustic generation. This implicit conflation of discriminative adequacy with generative viability underpins the \textbf{``unified'' discrete token} paradigm: the assumption that extreme compression, so long as it maintains
low WER, will not compromise synthesis quality.

\input{Image_set/methodology}

\section{The Methodological Bottleneck: Fixed-Stride Compression}
\label{sec:bottleneck}

To empirically validate or refute the ``unified token'' assumption, one must push semantic tokens to extreme compression rates (ultra-low frequencies) to isolate the purely linguistic state. However, conducting a rigorous stress test reveals a severe methodological bottleneck in current SLM architectures: the reliance on rigid, \textbf{fixed-stride downsampling}.

Standard compression mechanisms applied to semantic encoders (e.g., strided convolutions or fixed-window pooling) operate uniformly across the time axis \cite{radford2023robust,baevski2020wav2vec}. This rigid approach fundamentally conflicts with the physical reality of human speech, which is inherently \textbf{asynchronous}~\cite{ladgfoged2001course}. Phonetic events and acoustic micro-dynamics vary drastically in duration; for instance, a transient plosive (like /p/ or /t/) spans only a few milliseconds, whereas a stressed vowel may extend over hundreds of milliseconds~\cite{ladgfoged2001course,stevens2000acoustic}. 

When fixed-stride compression is aggressively scaled to achieve ultra-low frame rates, it acts as a blind temporal filter. It arbitrarily truncates and fractures these underlying phonetic boundaries, forcefully merging distinct acoustic events or splitting single phonemes across multiple discrete tokens~\cite{lu2016segmental,dong2020cif}. Consequently, this temporal misalignment artificially corrupts the semantic integrity of the sequence, causing the downstream WER to explode before the generative limits of the tokens can be fairly evaluated. 

This architectural flaw creates a confounding variable: when TTS generation fails at low frame rates, it is often misattributed to the corrupted alignment (the exploded WER) rather than the intrinsic generative deficiency of the purely semantic tokens~\cite{gong2025xy,cheng2025dycast}. Therefore, to truly isolate a ``pure'' semantic representation without sacrificing comprehension accuracy, we require a paradigm shift in tokenization. We must abandon fixed-stride constraints in favor of a \textbf{dynamic compression mechanism}---one that intelligently integrates information and aligns token emissions strictly with the asynchronous semantic boundaries of the continuous speech signal~\cite{dong2020cif,gao2022paraformer}.

\section{Methodology: Dynamic Compression Tokenizer}
\label{sec:methodology}

To explicitly isolate pure semantic representations at ultra-low frame rates while overcoming the arbitrary truncation of fixed-stride mechanisms, we propose a novel dynamic tokenization architecture. The rationale behind this design is deeply rooted in the physical reality of human speech: acoustic transitions are inherently asynchronous. Drawing inspiration from the linguistic theory of distinctive features~\cite{liu1996landmark,stevens2002toward,he2019ctc,zhang2025auto,zhang2025pre,zhang2025speecht} and dynamic sequence merging techniques in other domains~\cite{dong2020cif,gao2022paraformer,baade2025syllablelm}, our architecture is meticulously designed to align token emissions with actual semantic boundaries rather than rigid temporal grids. 

% As illustrated in Figure \ref{fig:architecture}, our pipeline realizes this through four synergistic components: (1) a pre-trained continuous semantic encoder (Whisper) to extract robust linguistic latents; (2) a Dynamic Merge Module to soft-accumulate these continuous features according to their distinctive acoustic transitions; (3) a Finite Scalar Quantization bottleneck to strictly enforce extreme discrete compression; and (4) a multi-objective decoding framework to independently probe the semantic integrity and generative viability of the resulting tokens.

As illustrated in Figure \ref{fig:architecture}, our pipeline
realizes this through four components, each motivated by a specific requirement identified in the preceding analysis. First, we employ a pre-trained Whisper encoder as the continuous semantic backbone, leveraging its established linguistic robustness to avoid introducing encoder quality as a confounding variable. Second, a Dynamic Merge Module soft-accumulates these continuous features according to their acoustic transitions, directly addressing the fixed-stride boundary corruption identified in Section \ref{sec:bottleneck}. Third, a Finite Scalar Quantization bottleneck enforces strict discrete compression, providing the informational constraint necessary to push tokens toward the ultra-low-rate regime under test. Finally, a multi-objective decoding framework branches into independent discriminative and generative pathways, enabling the Dual-Probing Protocol that evaluates semantic integrity and generative viability from the same token sequence.

\subsection{Information Weight Prediction and Soft-Accumulation}
\label{subsec:weight_prediction}
Given an input Log-Mel Spectrogram, the semantic encoder extracts a sequence of continuous acoustic features $H = \{h_1, h_2, \dots, h_{T}\} \in \mathbb{R}^{T \times D}$, where $T$ is the number of acoustic frames. Instead of rigidly downsampling $H$ by a fixed temporal window, our Dynamic Merge Module predicts a frame-level information weight $\alpha_t \in (0, 1)$ for each frame $h_t$. 

A 1D convolutional layer followed by a linear projection and a sigmoid activation is employed to estimate these weights. To enhance sparsity and suppress acoustic noise, we apply a smoothing factor $\lambda_{s}$ and a noise threshold $\lambda_{n}$:
\begin{equation}
    \alpha_t = \text{ReLU}(\sigma(\text{Conv1D}(H)_t \cdot \mathbf{W} + b) \cdot \lambda_{s} - \lambda_{n})
\end{equation}
The sequence of weights $\alpha = \{\alpha_1, \dots, \alpha_T\}$ represents the semantic boundary probabilities. In standard dynamic formulations, a new token boundary is dynamically triggered whenever the accumulated weight $S_t = \sum_{\tau=1}^t \alpha_\tau$ exceeds an integer threshold (typically $1.0$). 

\subsection{Macroscopic Fixed-Ratio with Microscopic Dynamic Alignment}
\label{subsec:fixed_ratio}
Conventional mechanisms determine the final sequence length based dynamically on the latent semantic content, rendering the overall compression rate unpredictable. However, to empirically stress-test the ``unified token'' assumption, we must evaluate the tokens at a strictly controlled, ultra-low frame rate. To bridge this gap, we introduce an \textbf{Information Scaling Paradigm}.

We impose a strict, predefined global compression ratio $R$ (e.g., $R=8$ converts 50Hz features to 6.25Hz). For an input sequence of length $T$, we explicitly compute the exact target compressed length $N = \max(1, \lfloor T / R \rceil)$. Prior to the integrate-and-fire operation, we globally scale the predicted information weights $\alpha_t$ such that their total sum strictly equals the target length $N$:
\begin{equation}
    \hat{\alpha}_t = \alpha_t \cdot \frac{N}{\sum_{i=1}^{T} \alpha_i}
\end{equation}
This scaling constraint (\textbf{Macroscopic Fixed-Ratio}) forces the module to compress the entire utterance into exactly $N$ tokens. Crucially, the network retains complete freedom over the distribution of $\hat{\alpha}_t$ (\textbf{Microscopic Dynamic Alignment}). It learns to reallocate its constrained ``token budget,'' allocating higher weights to dense semantic transitions and lower weights to steady-state acoustics. The continuous frames $h_t$ are then softly aggregated into a sequence of dynamically merged continuous tokens $C = \{c_1, \dots, c_N\}$.

\subsection{Quantization and Dual-Decoder Architecture}
\label{subsec:quantization_decoding}
To enforce a rigorous discrete informational bottleneck, the merged sequence $C$ is passed through an FSQ Quantizer, mapping each $c_j$ to a discrete codebook vector to obtain the final semantic tokens $Z = \{z_1, z_2, \dots, z_N\}$.

To expose the generative limitations of these highly compressed tokens, our architecture bifurcates into a dual-decoding pathway. 
First, $Z$ is fed into a \textbf{Whisper Decoder} and a \textbf{CTC Layer}~\cite{watanabe2017hybrid} for autoregressive generation and alignment, ensuring the tokens maintain absolute semantic integrity (measured by WER). 
Second, $Z$ is optionally passed to a lightweight \textbf{Recon Decoder}. To completely eliminate temporal misalignment as a confounding variable in generation, we utilize the cumulative weights $\hat{S}_t = \sum_{\tau=1}^t \hat{\alpha}_\tau$ to perform an Oracle Target Length Upsampling. The compressed sequence $Z$ is mapped exactly back to the original temporal resolution $T$ via $u_t = z_{\lfloor \hat{S}_t \rfloor}$, which explicitly aligns the discrete tokens to their physical acoustic boundaries before acoustic reconstruction.

\input{Image_set/evaluation}
\subsection{Overall Optimization Objectives}
\label{subsec:optimization}
The entire architecture is optimized end-to-end using a multi-task learning framework. The total loss $\mathcal{L}_{total}$ comprises fundamental objectives for semantic alignment and compression, along with an optional generative probe:
\begin{equation}
\begin{split}
    \mathcal{L}_{total} &= \mathcal{L}_{CTC} + \mathcal{L}_{Attn} \\
    &\quad + \lambda_{qua}\mathcal{L}_{Qua} + \lambda_{recon}\mathcal{L}_{Recon}
\end{split}
\end{equation}

\textbf{Semantic Understanding Losses.} To guarantee that the ultra-low frequency tokens capture pure linguistic content, we apply a Connectionist Temporal Classification loss ($\mathcal{L}_{CTC}$) on the latent projections and an Attention-based cross-entropy loss ($\mathcal{L}_{Attn}$) from the Whisper Decoder. Both are strictly supervised by the ground-truth target text.

\textbf{Quantity Loss.} To guide the Dynamic Merge Module in predicting reasonable semantic boundaries before the forced scaling is applied, we penalize the $L_1$ distance between the predicted unscaled token sum and the exact target compressed length $N$:

\begin{equation}
    \mathcal{L}_{Qua} = \frac{1}{\sum N} \sum \left| \sum_{t=1}^T \alpha_t - N \right|
\end{equation}

\textbf{Acoustic Reconstruction Loss.} As an \textit{optional} generative probe to explicitly test the ``unified token'' hypothesis, we train the Recon Decoder to reconstruct the original Log-Mel Spectrogram from the upsampled discrete tokens. This process is supervised by a Mean Squared Error loss ($\mathcal{L}_{Recon}$) applied directly to the continuous acoustic space.

\section{Evaluation Framework: The Dual-Probing Protocol}
\label{sec:evaluation}

The dynamic tokenizer developed in Section \ref{sec:methodology} provides the necessary control variable: it isolates pure semantic states via extreme compression without introducing the confounding alignment errors typical of fixed-stride methods. To test whether these isolated tokens are sufficient for both understanding and generation, we propose a Dual-Probing Protocol that evaluates the exact same token sequence $Z$ through two independent pathways.

\textbf{Discriminative Understanding Probe.} To validate semantic integrity beyond WER, we formulate audio understanding as a multiple-choice AQA classification task~\cite{sakshi2024mmau}. As illustrated in Figure~\ref{fig:downstream}, the discrete FSQ codes are projected into a frozen LLM's embedding space via a trainable Audio Projector. By freezing the LLM backbone, any downstream accuracy is exclusively attributable to the informational density of the tokens themselves, not the reasoning capacity of the language model.

\textbf{Generative Probe.} To stress-test generative viability, we employ a Flow Matching decoder~\cite{lipman2022flow} conditioned on the oracle upsampled sequence $U$ (Section \ref{subsec:quantization_decoding}), which guarantees perfect temporal alignment. This forces the generative model to rely entirely on the micro-temporal dynamics \textit{within} the tokens. If the tokens excel in the Discriminative Probe but fail here, it constitutes direct evidence that extreme semantic compression eradicates the continuous gradients required for ODE-based synthesis. Full architectural details and training objectives for both probes are provided in Appendix~\ref{app:dual_probing}.

\section{Experiments}
\label{sec:experiments}

\subsection{Experimental Setup}
\label{subsec:exp_setup}

\textbf{Datasets.} We train our dynamic tokenizer using a
large-scale, multi-lingual corpus comprising thousands of hours of
speech from LibriSpeech~\cite{panayotov2015librispeech},
GigaSpeech~\cite{chen2021gigaspeech}, and
Aishell~\cite{fu2021aishell}, among other open-source datasets,
ensuring robust phonetic coverage across English and Mandarin. For
evaluation, we report CER on two standard Mandarin test sets from
WenetSpeech~\cite{zhang2022wenetspeech}: \textbf{Test\_Net} (23
hours, internet speech, matched domain) and \textbf{Test\_Meeting}
(15 hours, real meeting recordings, mismatched domain with far-field
conditions).

\textbf{Model and Compression.} Our tokenizer backbone follows a
Whisper-style Transformer architecture. The input 128-dimensional
log-Mel spectrograms are first downsampled to 50Hz. We enforce a
compression ratio of $R{=}10$ within the Dynamic Merge Module,
yielding an ultra-low frame rate of 5.0Hz. The merged
representations are quantized via FSQ with 7 dimensions and 4 levels
per dimension (codebook size $4^7{=}16{,}384$). The model is
optimized end-to-end with Adam (peak LR $2{\times}10^{-5}$, 12k
warmup steps) using the multi-task objective in Section
\ref{subsec:optimization}. Full architectural specifications and
hyperparameter configurations are detailed in
Appendix~\ref{app:exp_config}.

\subsection{Discriminative Probe: Establishing the Semantic Upper Bound}
\label{subsec:exp_understanding}

The primary prerequisite for testing our ``ODE Flatline'' hypothesis is obtaining a sequence of tokens that is extremely compressed yet undeniably semantically rich. If the tokens used for our generative probe are semantically deficient, any subsequent acoustic collapse could be trivially dismissed as an artifact of information loss rather than a fundamental representational mismatch. We evaluate this semantic sufficiency through two dimensions: Character Error Rate (CER) on the WenetSpeech corpus (Table \ref{tab:understanding_results}) and complex language reasoning via the AVQA task (Table \ref{tab:avqa_results}).

\textbf{The Confounding Variable of Fixed-Stride.} As shown in Table \ref{tab:understanding_results} and Table \ref{tab:avqa_results}, scaling standard fixed-stride downsampling to an ultra-low 5.5Hz severely truncates necessary semantic information (CERs exploding to 29.80\% and 31.50\%). Consequently, its downstream AVQA accuracy collapses to 0.5777. This confirms that fixed-stride truncation inherently corrupts the linguistic state, rendering it an invalid control variable for testing generative limits at extreme compression.

\textbf{Isolating the Pure Semantic State.} In contrast, our Dynamic Compression Tokenizer completely bypasses this methodological bottleneck. To prove these tokens possess sufficient informational depth for generation, we compare them against established high-framerate SOTA tokenizers (e.g., WavTokenizer at $\sim$50Hz) used in standard generative SLMs. 

Operating at an ultra-low frame rate of \textbf{5.0Hz}, our dynamic tokens (w/ Recon) achieve a remarkable AVQA accuracy of \textbf{0.7139}, eclipsing the reasoning capabilities of standard 50-75Hz representations (0.65-0.67) and recent syllable-based models like SYLLABLELM (0.5526). \textbf{This comparison is not to claim a new state-of-the-art tokenizer}, but to establish a rigorous empirical baseline: our 5.0Hz tokens are unequivocally saturated with dense semantic and reasoning information.

\textbf{The Paradoxical Trap.} The most profound insight emerges from our ablation study. When the dynamic tokenizer is trained strictly without the acoustic reconstruction probe loss ($\mathcal{L}_{Recon}$), its downstream AVQA accuracy peaks at \textbf{0.7246} (Dynamic FSQ, Pure Semantic). This sets up a profound paradox. Relieved of the burden to preserve continuous acoustic micro-dynamics, the tokens' semantic categorization capability reaches its absolute upper bound. 
\input{table/main_result} 
We have successfully isolated a discrete representation that exceeds the semantic capacity of current generative benchmarks, yet it has been systematically stripped of acoustic constraints. Having established these 5.0Hz tokens as flawless, ultra-dense vehicles for language comprehension, we have completely eliminated ``information scarcity'' as a potential excuse. We now push these highly capable semantic tokens into the Generative Flow Matching Probe to expose the true nature of the unified token illusion.

\input{table/avqa_result} 

\subsection{Generative Probe: Exposing the ODE Flatline}
\label{subsec:exp_generation}
\input{Image_set/mel_analysis}

Having established in Tables~\ref{tab:understanding_results} and \ref{tab:avqa_results} that the ultra-low-rate dynamic tokens remain semantically competent, we now turn to the critical question of generation. The central purpose of the Generative Probe is not to evaluate reconstruction quality in isolation, but to determine whether a representation that is demonstrably sufficient for discriminative understanding can also support continuous acoustic synthesis. This distinction is essential. If generation fails only because token boundaries are corrupted by rigid downsampling, then the failure says little about the intrinsic adequacy of semantic tokens themselves. However, once dynamic compression removes this alignment confound and oracle upsampling restores the target temporal resolution, any remaining collapse must be attributed to the representational content of the tokens rather than to boundary mismatch.

Figure~\ref{fig:micro_blur_failure} makes this distinction explicit. The reconstructed utterance is nearly perfectly matched in duration to the source ($\texttt{duration\_ratio}=0.9995$), and the reconstructed log-Mel spectrogram still retains visible correspondence to the original at the level of the coarse spectral envelope ($\texttt{mel\_corr}=0.6285$). These two observations are important: they indicate that the token sequence has not undergone trivial semantic collapse, nor is the failure primarily explained by temporal misalignment. At a macroscopic level, the utterance remains structurally recognizable.

Yet this apparent success is precisely where the WER trap becomes visible. The decisive failure emerges not in the coarse envelope, but in the local dynamics. The reconstruction exhibits substantial deviations in the Mel domain ($\texttt{mel\_mae}=11.11$), and, more critically, severe errors in the temporal-difference domain ($\texttt{delta\_mae}=7.90$ and $\texttt{flux\_mae}=4.04$). These error maps show that the model fails to recover the rapid frame-to-frame spectral transitions that encode articulation, including the short-timescale modulations needed to realize phonetic boundaries and continuous acoustic motion. Perceptually, this manifests as catastrophic articulatory loss and severe acoustic muffling---the generated speech becomes effectively unintelligible. The degradation in these micro-dynamic metrics is so stark that it precludes the necessity of subjective Mean Opinion Score (MOS) evaluations to confirm the generative collapse. This systemic failure is rigorously validated across diverse utterance lengths and verified over the entire test set in Appendix~\ref{app:additional_recon_analysis} and Appendix~\ref{app:quantitative_recon_stats}. In other words, the compressed token sequence still carries enough information to preserve a semantically interpretable global structure, but it no longer provides the micro-dynamic gradients required to steer a continuous generative trajectory faithfully.

Taken together with the discriminative results---14.32/15.94 CER and 0.7139 AVQA accuracy at 5.0Hz, rising to 0.7246 under purely semantic training---these observations rule out information scarcity as an explanation for the generative failure. The tokens retain sufficient content for robust language understanding, yet the micro-dynamic gradients required by ODE-based synthesis have been irrevocably discarded. This confirms that low CER and generative adequacy are not interchangeable objectives: the very compression that sharpens semantic categorization strips away the continuous phonetic trajectories on which acoustic generation depends.
\vspace{-2pt}
\section{Conclusion}
\label{sec:conclusion}
\vspace{-2pt}
This work exposes a fundamental flaw in the prevailing assumption that a single discrete token, validated by low WER, can universally serve both speech understanding and generation. By developing a dynamic compression tokenizer that achieves ultra-low frame rates without corrupting semantic boundaries, we isolate pure semantic representations as a rigorous control variable. Our Dual-Probing Protocol reveals a stark divergence: at 5.0Hz, these tokens attain strong discriminative performance---surpassing standard high-framerate representations in complex audio reasoning---yet cause severe articulatory collapse when conditioning ODE-based generative models under oracle temporal alignment. This \emph{ODE Flatline} is not an engineering artifact but an information-theoretic consequence: the categorical abstraction rewarded by WER optimization is structurally orthogonal to the continuous micro-dynamic gradients demanded by acoustic synthesis. Our findings carry a direct implication: the pursuit of a unified, ultra-compressed speech token is fundamentally misguided. We advocate instead for explicitly decoupled representations---semantic tokens for understanding, acoustic tokens for generation---as the principled path toward truly capable speech language models.

\section*{Limitations}

Our generative probe employs a single synthesis paradigm, ODE-based Flow Matching, which is currently the dominant framework for discrete-token-conditioned speech generation. While alternative architectures such as autoregressive waveform models or GAN-based vocoders may exhibit different failure modes, the structural root cause we identify---the erasure of continuous micro-dynamic gradients under extreme discrete compression---is upstream of any particular decoder; it resides in the token representation itself, and is therefore expected to constrain any generation method that requires sub-token temporal detail. All evaluations are conducted on Mandarin (WenetSpeech), a tonal language whose lexical pitch contours impose stringent micro-dynamic demands. This arguably makes our experimental setting a strong test case rather than a narrow one: if the representational gap already manifests in a language where fine pitch variation is semantically critical, it is unlikely to vanish in phonetically simpler contexts. Finally, this work is diagnostic rather than constructive---we expose the incompatibility but do not propose a concrete decoupled architecture. However, precisely delineating the failure boundary is a necessary prerequisite for principled design, and we believe the Dual-Probing Protocol introduced here provides a reusable evaluation framework for future work on decoupled speech representations.

\bibliography{custom}

\appendix
\section{Detailed Experimental Configurations}
\label{app:exp_config}
\input{Image_set/otheranalysis}

To facilitate reproducibility, we detail the complete hyperparameter configurations used for training the Dynamic Compression Tokenizer, the quantization bottleneck, and the generative reconstruction probe. 

\textbf{Acoustic Features.} The input signals are processed into 128-dimensional log-Mel spectrograms, extracted using a 400-point FFT and a hop length of 160. The sampling rate is standardized to 16kHz. 

\textbf{Tokenizer Backbone.} Both the encoder and decoder utilize a robust Transformer architecture:
\begin{itemize}
    \item \textbf{Encoder:} 32 layers, 20 attention heads, 1280 hidden dimension, and 5120 linear units. The input layer uses a \texttt{conv1d2} module, which provides an initial $2\times$ temporal downsampling (resulting in 50Hz features).
    \item \textbf{Decoder:} 32 layers, 20 attention heads, 1280 hidden dimension, utilizing Whisper-style subword tokenization (vocabulary size: 51,866, including special tokens like \texttt{<sot>} and \texttt{<eot>}).
\end{itemize}

\textbf{Dynamic Merge Module.} The dynamic compression operates with the following hyperparameters to strictly enforce the ultra-low frame rate:
\begin{itemize}
    \item \textbf{Integration Threshold:} 1.0, with a smooth factor of 1.0 and a noise threshold of 0.0.
    \item \textbf{Custom Initialization:} The output bias of the dynamic predictor is initialized to -1.5. This negative bias ensures smaller initial information weights ($\alpha_t$), stabilizing the early stages of training under extreme compression constraints.
    \item \textbf{Train-Infer Consistency:} Scaling is strictly enforced during both training and inference to guarantee exactly aligned $N = \lfloor T/R \rceil$ token lengths.
\end{itemize}

\textbf{Quantization (FSQ).} The discrete bottleneck is applied immediately after the 32nd encoder layer. We employ Finite Scalar Quantization (FSQ) mapped to 7 dimensions, with 4 levels per dimension (i.e., $[4, 4, 4, 4, 4, 4, 4]$). Positional embeddings are injected post-quantization to retain global temporal awareness for the downstream probes.

\textbf{Generative Reconstruction Probe.} The lightweight continuous decoder used to verify acoustic collapse comprises a 3-layer 1D ResNet architecture. It takes the 1280-dimensional upsampled discrete tokens and reconstructs the 128-dimensional log-Mel spectrograms. The network utilizes Snake activations \cite{ziyin2020neural} and employs hierarchical transposed convolutions with strides of $[2, 2, 2]$ to achieve an $8\times$ upsampling back to the acoustic resolution.

\textbf{Optimization.} The model is optimized using Adam with a peak learning rate of $2 \times 10^{-5}$ and 12,000 warmup steps. We use dynamic batching (max 24,000 frames per batch) with a gradient accumulation step of 2 and gradient clipping capped at 5.0.

\section{Dual-Probing Protocol: Architectural Details}
\label{app:dual_probing}

This appendix provides the complete architectural and training details for the two evaluation pathways introduced in Section~\ref{sec:evaluation}.

\subsection{Discriminative Understanding Probe via Frozen LLMs}
\label{app:understanding_probe}
Relying solely on WER is insufficient for modern SLMs; therefore, drawing inspiration from recent advances in discrete token projection techniques and LLM-based audio evaluation paradigms~\cite{deshmukh2023pengi}, we formulate audio understanding as a multiple-choice AQA classification task~\cite{sakshi2024mmau}. To strictly isolate the representational quality of $Z$ from the intrinsic reasoning capacity of the LLM, we employ a completely frozen language model backbone.

The discrete FSQ codes are mapped into the continuous semantic space via a trainable Audio Projector:
\begin{equation}
    E_{audio} = \text{Network}(\text{Embed}(Z)) \in \mathbb{R}^{N \times d_{llm}}
\end{equation}
The projected embeddings $E_{audio}$ are concatenated with the text embeddings of the question and its options. By freezing the LLM and only training the projector and classification head, any success in downstream accuracy is exclusively attributable to the informational density of the underlying tokenizer. This probe is explicitly designed to confirm that our dynamic compression preserves deep linguistic reasoning capabilities, establishing the tokens as highly effective for comprehension tasks.

\subsection{Generative Probe via Continuous Flow Matching}
\label{app:generative_probe_fm}

We employ a Flow Matching decoder, the current standard for Whisper-style discrete generation. As hypothesized, high-fidelity acoustic generation via FM requires the conditioning signal to provide temporally dynamic gradients to the ODE solver. To test this without the confounding variable of temporal misalignment, we condition the FM model on the oracle upsampled sequence $U \in \mathbb{R}^{T \times D}$ (as defined in Section \ref{subsec:quantization_decoding}). The generative process is formulated as an ODE, $\frac{d x_t}{dt} = v_\theta(x_t, t, U)$, where $v_\theta$ is trained to match the target vector field between a Gaussian prior $x_0$ and the continuous log-Mel spectrogram $x_1$:

\small
\begin{equation}
    \mathcal{L}_{FM} = \mathbb{E}_{t, x_1, x_0} \left[ \left\| v_\theta(x_t, t, U) - (x_1 - x_0) \right\|_2^2 \right]
\end{equation}
\normalsize

By explicitly feeding the FM decoder with $U$---which guarantees perfect physical acoustic boundaries---we force the generative model to rely entirely on the micro-temporal dynamics \textit{within} the tokens. If the tokens excel in the Discriminative Probe but fail here, resulting in severe articulation blur, it serves as definitive empirical proof of our hypothesis.

\section{Additional Qualitative Evidence for the ODE Flatline}
\label{app:additional_recon_analysis}

Figure~\ref{fig:appendix_recon_gallery} extends the main-text qualitative evidence by presenting reconstruction results across short, medium, and long utterances, as well as an extreme collapse case. The recurrence of the same failure pattern across substantially different durations confirms that the phenomenon is systematic rather than sample-specific. Across all 495 reconstruction cases, the duration ratio is tightly concentrated around 1.0 (median 0.9978, interquartile range 0.9965--0.9989), indicating that oracle upsampling successfully eliminates temporal mismatch as a confounding factor. The failure therefore cannot be attributed to duration error or token-frame misalignment.

In the short-, medium-, and long-duration examples, the reconstructed log-Mel spectrograms retain a rough correspondence to the source at the level of the coarse spectral envelope---broad energy placement, phrase-level segmentation, and approximate frequency-band occupancy remain visible. This confirms that the compressed tokens still encode sufficient information to preserve a global semantic scaffold, consistent with their strong CER and AVQA performance reported in the main text. However, the dominant deviations are consistently concentrated in the fine-grained structures that reflect rapid spectral transitions and short-lived articulatory events. The temporal-difference and spectral-flux error maps reveal that the largest errors arise precisely in the frame-to-frame change patterns that encode phonetic boundaries and continuous acoustic motion. Notably, even the short-duration example ($\sim$2 seconds) exhibits the same qualitative signature, ruling out the explanation that micro-dynamic blur is merely a byproduct of long-horizon error accumulation. The bottleneck lies in the representational content of the token stream itself.

The extreme collapse case in the final row serves a complementary role: it demonstrates that once the conditioning signal becomes sufficiently impoverished, even the coarse spectral scaffold breaks down entirely. Together, these examples delineate a continuum of failure severity---from semantically scaffolded but articulatorily blurred reconstructions to near-total spectral collapse---providing further evidence that the gap between discriminative adequacy and generative adequacy is a persistent structural property of ultra-low-rate semantic tokens, not an artifact of any single experimental condition.

\section{Quantitative Reconstruction Statistics}
\input{table/mel_table}
\label{app:quantitative_recon_stats}

To complement the qualitative examples in Figures~\ref{fig:micro_blur_failure} and~\ref{fig:appendix_recon_gallery}, Table~\ref{tab:appendix_recon_metrics} summarizes the reconstruction statistics over all 495 evaluated utterances. Since the dynamic tokenizer removes boundary corruption introduced by fixed-stride compression, and oracle duration alignment eliminates gross temporal mismatch during upsampling, the remaining errors can be attributed more directly to representational insufficiency rather than architectural confounds.

The most revealing pattern is the sharp asymmetry between temporal preservation and spectral fidelity. The Duration Ratio is tightly concentrated near 1.0 (median 0.9978, IQR 0.9965--0.9989), confirming that reconstructed utterances preserve nearly correct temporal extent in the vast majority of cases. The decoder is not failing because the utterance is globally misaligned---this rules out the familiar alignment pathology of low-rate tokenization as the primary explanation.

However, this temporal fidelity does not translate into acoustic intelligibility. The median Mel correlation is only 0.471, indicating at best a partial coarse resemblance to the source. More critically, the delta-domain errors remain persistently high: the median Delta-Mel MAE is 6.41 and the median Flux MAE is 3.28. These metrics probe frame-to-frame spectral evolution rather than static spectral occupancy, and their magnitude confirms that the dominant failure lies in the rapid local transitions encoding articulation---precisely the micro-dynamic gradients that a continuous generative decoder requires to steer its acoustic trajectory.

This asymmetry is exactly what one would predict if the token stream retains categorical semantic states but has lost the continuous micro-dynamic cues required for generation. A low-rate semantic tokenizer can preserve utterance-level structure, broad phonetic regions, and a coarse spectral envelope---information that is highly useful for discriminative objectives, consistent with the strong CER and AVQA results in the main text. But continuous synthesis additionally requires knowledge of \emph{how} the spectrum should locally evolve between articulatory configurations, and this is precisely where the representation fails. The quantitative footprint thus mirrors the qualitative evidence: global timing survives, local dynamics do not. This dissociation provides aggregate statistical confirmation that the failure of ultra-low-rate semantic tokens is not one of generic quality degradation, but a structural mismatch between discriminative sufficiency and generative adequacy.

\end{document}

%% file: Image_set/concept.tex
\begin{figure}[t]
    \centering
    \includegraphics[width=0.5\textwidth]{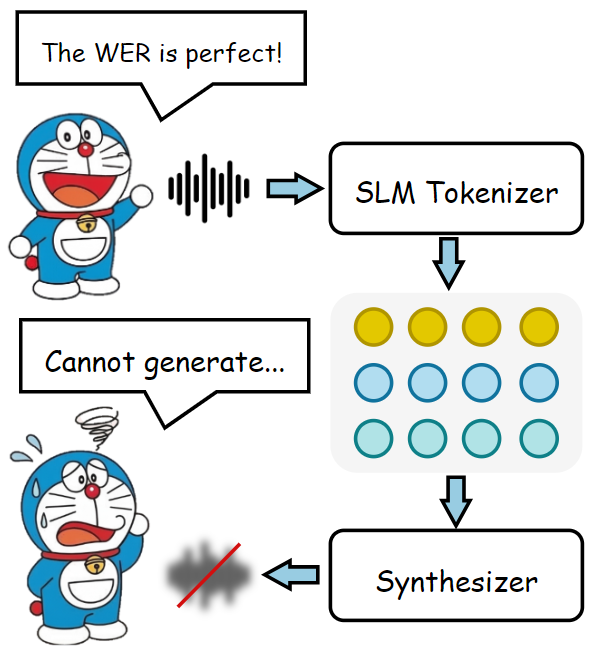}
\caption{A conceptual illustration of the \textbf{WER Trap} It highlights a fundamental paradox in Speech Language Models: discrete tokens that achieve perfect Word Error Rate (WER) in semantic comprehension (top) inherently discard the fine-grained acoustic details required for generation, resulting in synthesis failure (bottom)}
    \label{fig:concept}
\end{figure}

%% file: Image_set/methodology.tex
\begin{figure*}[!t]
    \centering
    \includegraphics[width=\textwidth]{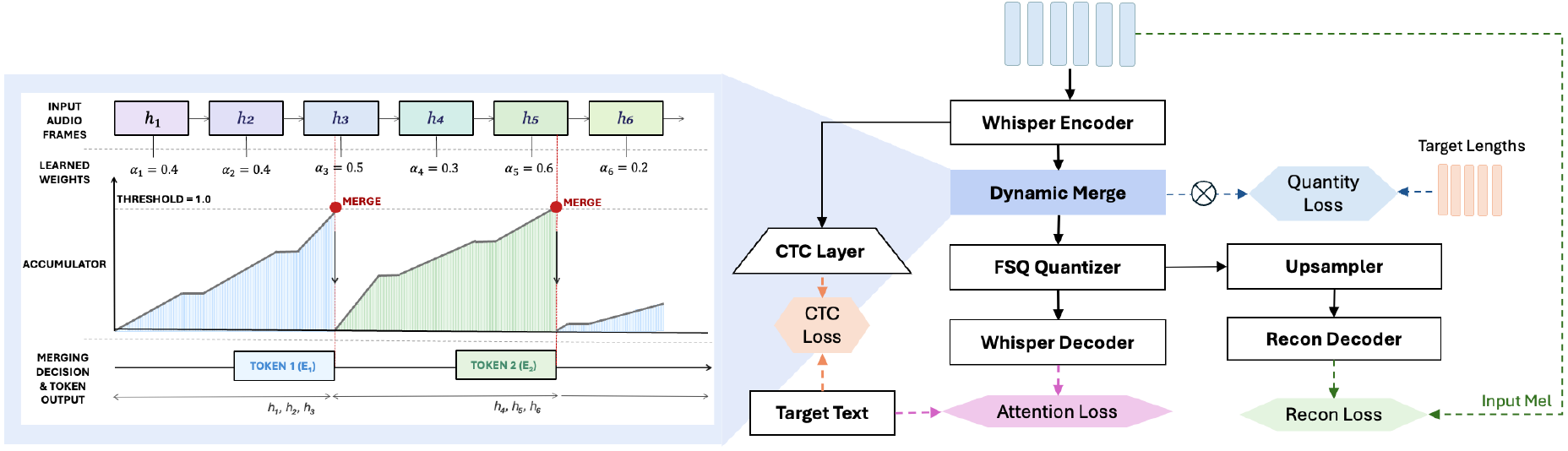}
    \caption{The Dynamic Compression Tokenizer. \textbf{Left:} The soft-accumulation mechanism of the Dynamic Merge Module. Learned frame weights $\alpha_t$ are progressively summed until the threshold $\Theta = 1.0$ is reached, triggering a token boundary. Frames $h_1, h_2, h_3$ are soft-aggregated into Token 1 ($E_1$), and $h_4, h_5, h_6$ into Token 2 ($E_2$). \textbf{Right:} The complete dual-probing architecture.}
    \label{fig:architecture}
\end{figure*}

%% file: Image_set/evaluation.tex
\begin{figure}[t]
    \centering
    \includegraphics[width=0.5\textwidth]{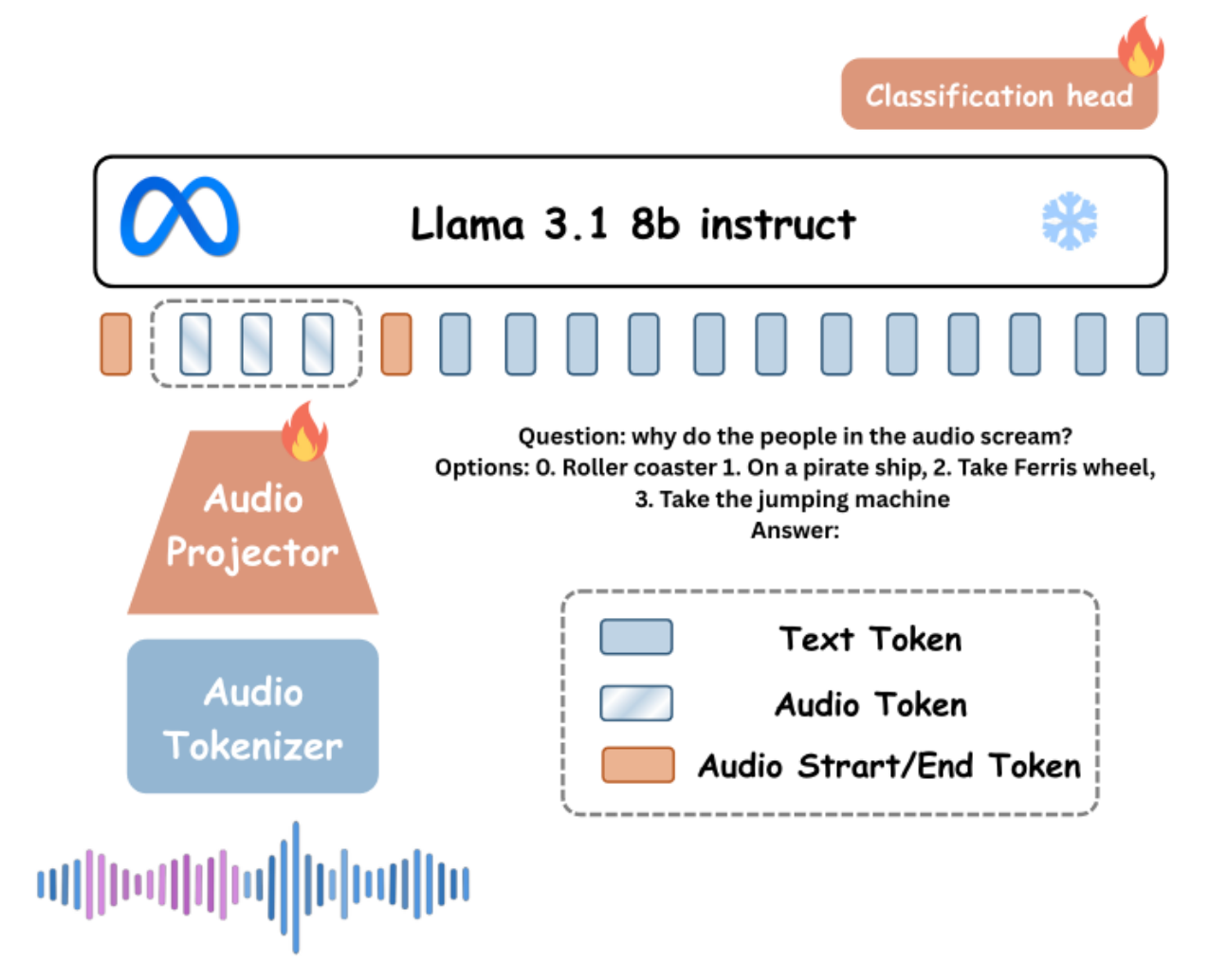}
\caption{Downstream evaluation framework. Discrete VQ codes from the audio tokenizer are mapped to language model embeddings via a trainable Audio Projector. The Llama 3.1 8B model (frozen) processes audio tokens with textual questions for classification. Only the Audio Projector and classification head are trainable, isolating tokenization quality effects.}
    \label{fig:downstream}
\end{figure}

%% file: table/main_result.tex
\begin{table}[t]
\centering
\small
\caption{Character Error Rate (CER \%) comparison on WenetSpeech test sets. The results highlight the catastrophic semantic degradation of fixed-stride compression at ultra-low frequencies ($\sim$5.5Hz) and the robust semantic preservation of our Dynamic Compression Tokenizer. $^\dagger$Trained without the acoustic reconstruction probe loss ($\mathcal{L}_{Recon}$), isolating the purely linguistic state.}
\label{tab:understanding_results}
\resizebox{\linewidth}{!}{
\begin{tabular}{lccc}
\toprule
\textbf{Model} & \textbf{Rate} & \textbf{Test-Net} & \textbf{Test-Meeting} \\
\midrule
\multicolumn{4}{l}{\textit{Continuous Reference (Uncompressed)}} \\
Whisper v3 Baseline & 50Hz & 9.68 & 18.54 \\
\midrule
\multicolumn{4}{l}{\textit{Fixed-Stride Compression}} \\
Whisper + FSQ & 12.5Hz & 18.66 & 20.46 \\
Whisper + FSQ & 5.5Hz & 29.80 & 31.50 \\
\midrule
\multicolumn{4}{l}{\textit{Dynamic Compression (Ours)}} \\
Dynamic FSQ & 6.0Hz & 14.47 & 16.14 \\
Dynamic FSQ (w/ Recon) & 5.0Hz & 14.32 & 15.94 \\
Dynamic FSQ (Pure Semantic)$^\dagger$ & 5.0Hz & \textbf{11.98} & \textbf{12.50} \\
Dynamic FSQ & 4.0Hz & 15.61 & 17.44 \\
\bottomrule
\end{tabular}
}
\end{table}

%% file: table/avqa_result.tex
\begin{table}[t]
\centering
\small
\caption{Downstream LLM Comprehension Performance (AVQA Accuracy). Evaluated using a frozen LLaMA 3.1 8B backbone. Our 5.0Hz dynamic tokens set a rigorous empirical baseline for semantic saturation. \textbf{$^\dagger$Trained without the acoustic reconstruction probe loss ($\mathcal{L}_{Recon}$).}}
\label{tab:avqa_results}
\resizebox{\linewidth}{!}{
\begin{tabular}{llc}
\toprule
\textbf{Model} & \textbf{Approx. Rate} & \textbf{AVQA Acc.} \\
\midrule
\multicolumn{3}{l}{\textit{Established SOTA Baselines}} \\
SYLLABLELM & $\sim$4-5Hz & 0.5526 \\
Speech Tokenizer & 50Hz & 0.5839 \\
DAC Tokenizer & 75-100Hz & 0.6561 \\
WavTokenizer & 40/75Hz & 0.6732 \\
\midrule
\multicolumn{3}{l}{\textit{Fixed-Stride Compression}} \\
Fixed-Stride FSQ & $\sim$5.5Hz & 0.5777 \\
\midrule
\multicolumn{3}{l}{\textit{Dynamic Compression (Ours)}} \\
Dynamic FSQ & 6.0Hz & 0.7015 \\
Dynamic FSQ (w/ Recon) & 5.0Hz & 0.7139 \\
Dynamic FSQ (Pure Semantic)$^\dagger$ & 5.0Hz & \textbf{0.7246} \\
Dynamic FSQ & 4.0Hz & 0.6526 \\
\bottomrule
\end{tabular}
}
\end{table}

%% file: Image_set/mel_analysis.tex
\begin{figure*}[t]
  \centering
  \includegraphics[width=\textwidth]{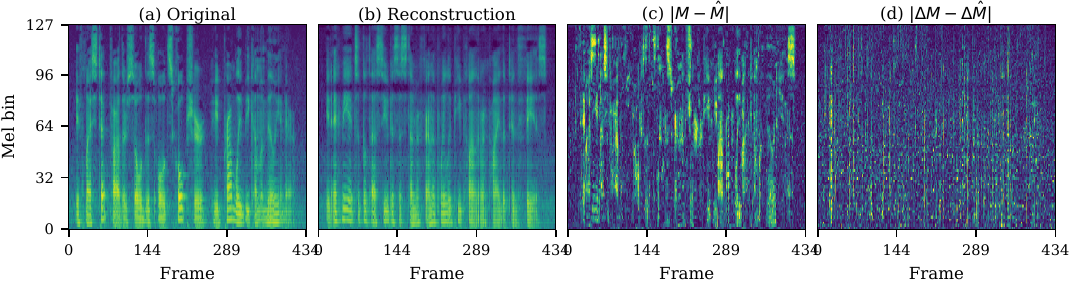}
  \caption{A representative reconstruction failure case from the dynamic tokenizer. Panels (a) and (b) show that the reconstructed log-Mel spectrogram roughly preserves the coarse semantic envelope and utterance duration, but panels (c) and (d) reveal substantial frame-wise and temporal-difference errors. This pattern indicates that ultra-low-rate semantic tokens retain enough information for coarse content recovery while failing to preserve the micro-dynamics required for clear articulation.}
  \vspace{-14pt}
  \label{fig:micro_blur_failure}
\end{figure*}

%% file: Image_set/otheranalysis.tex
\begin{figure*}[t]
  \centering
  \includegraphics[width=0.98\textwidth]{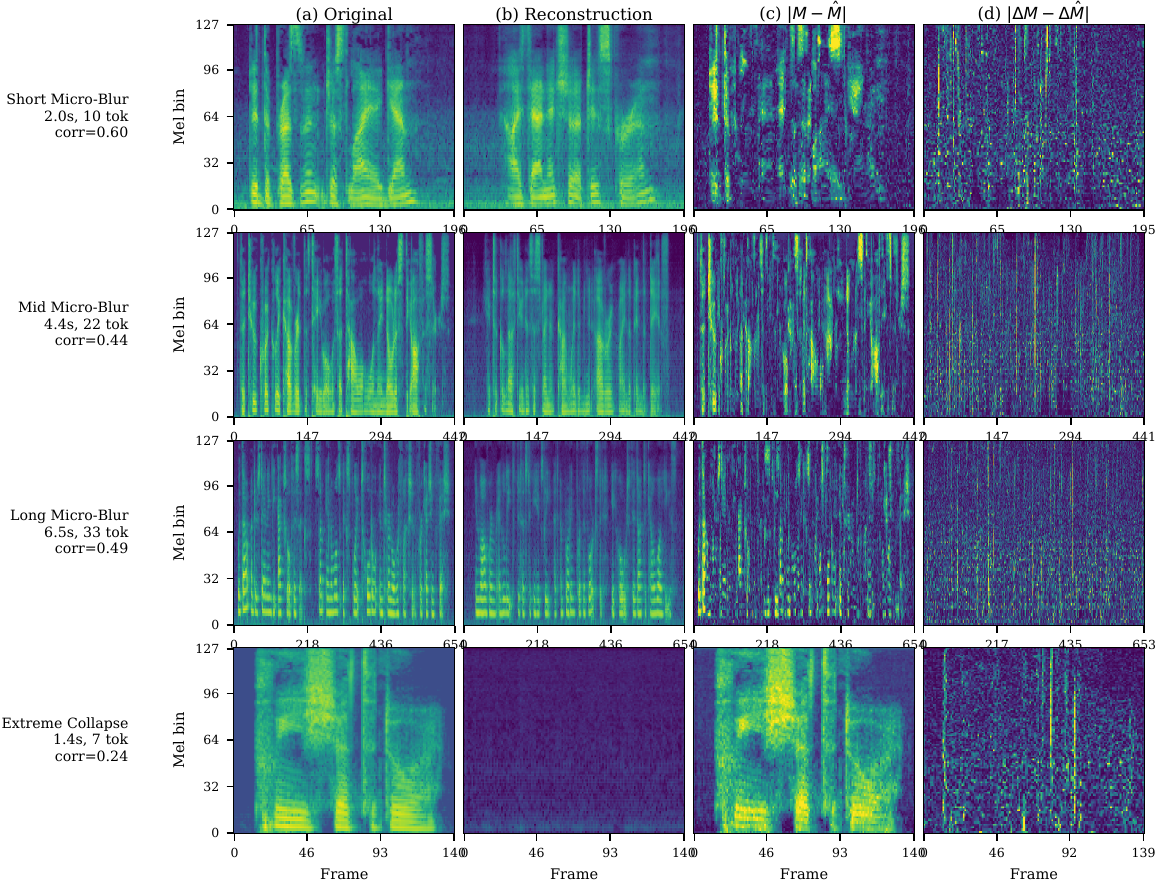}
  \caption{Additional reconstruction failures from the dynamic tokenizer under oracle duration alignment. Rows correspond to short-, medium-, and long-duration utterances, plus an extreme collapse case. Across diverse utterance lengths, the reconstructed spectrograms retain gross utterance extent but still exhibit large absolute and temporal-difference deviations, indicating that the generative deficiency is systematic rather than confined to a single sample.}
  \label{fig:appendix_recon_gallery}
\end{figure*}

%% file: table/mel_table.tex
\begin{table}[t]
  \centering
  \small
  \begin{tabular}{lcc}
    \toprule
    Metric & Median & IQR \\
    \midrule
    Mel MAE $\downarrow$ & 14.99 & 13.59--16.68 \\
    Mel Corr $\uparrow$ & 0.471 & 0.390--0.543 \\
    Delta-Mel MAE $\downarrow$ & 6.41 & 6.05--6.80 \\
    Flux MAE $\downarrow$ & 3.28 & 3.09--3.48 \\
    Duration Ratio $\uparrow$ & 0.9978 & 0.9965--0.9989 \\
    \bottomrule
  \end{tabular}
  \caption{Distribution of reconstruction metrics over all 495 dynamic-tokenizer reconstruction cases. We report the median and interquartile range (IQR) to avoid over-emphasizing a small number of extreme collapse samples. Duration Ratio close to 1.0 indicates that the reconstructed utterances largely preserve their global temporal extent, whereas the persistent Mel- and delta-domain errors reveal substantial spectral and micro-dynamic mismatch.}
  \label{tab:appendix_recon_metrics}
\end{table}